\documentclass[prd,twocolumn,amsmath,nobibnotes,nofootinbib,superscriptaddress,a4paper,floatfix]{revtex4} 

\usepackage{bm} 
\usepackage{graphicx} 

\def\ba{\begin{eqnarray}}
\def\ea{\end{eqnarray}}
\def\be{\begin{equation}}
\def\ee{\end{equation}}
\def\({\left(}
\def\){\right)}
\def\[{\left[}
\def\]{\right]}
\def\<{\left<}
\def\>{\right>}

\newcommand{\rar}{\rightarrow}

\newcommand{\labeq}[1] {\label{eq:#1}}
\newcommand{\refeq}[1] {(\ref{eq:#1})}
\newcommand{\labfig}[1] {\label{fig:#1}}
\newcommand{\reffig}[1] {\ref{fig:#1}}
\newcommand{\labsec}[1] {\label{sec:#1}}
\newcommand{\refsec}[1] {\ref{sec:#1}}

\newcommand{\ri} {\ensuremath{r_\text{i}}}
\newcommand{\rf} {\ensuremath{r_\text{f}}}

\begin{document}

\title{Path Integral for Stochastic Inflation: Non-Perturbative
  Volume Weighting, Complex Histories, Initial Conditions and the End
  of Inflation}
\date{March 11, 2010}
\author{Steven Gratton}
\email{stg20@cam.ac.uk}
\affiliation{Kavli Institute for Cosmology and Institute of Astronomy, 
Madingley Road, Cambridge, CB3 0HA, UK}

\begin{abstract}
In this paper we present a path integral formulation of stochastic
inflation, in which volume weighting can easily be implemented.  With
an in-depth study of inflation in a quartic potential, we investigate
how the inflaton evolves and how inflation typically ends both with
and without volume weighting.  Perhaps unexpectedly, complex histories
sometimes emerge with volume weighting.  The reward for this excursion
into the complex plane is an insight into how volume-weighted
inflation both loses memory of initial conditions and ends via
slow-roll.  The slow-roll end of inflation mitigates certain
``Youngness Paradox''-type criticisms of the volume-weighted paradigm.
Thus it is perhaps time to rehabilitate proper time volume weighting
as a viable measure for answering at least some interesting cosmological
questions.

\end{abstract}

\maketitle

\section{\labsec{intro}Introduction}

Inflation driven by the potential energy of some effective scalar
field \cite{Guth:1980zm,Linde:1981mu,Albrecht:1982wi} has become a
common explanation of the starting state of the radiation-dominated
hot big bang model.  A key reason for its acceptance is that small
quantum fluctuations during the last 60 or so efolds of inflation can
develop into almost scale invariant curvature perturbations
\cite{Hawking:1982cz,Guth:1982ec,Bardeen:1983qw} like those that we
see today in the cosmic microwave background fluctuations
\cite{Jarosik:2010iu}.  Couplings in the inflaton's potential have to
be chosen to be very small in order to get the amplitude of the
fluctuations suitably low.  However, fluctuations in the scalar field
increase as the background energy density increases, so in certain
circumstances the fluctuations might have a significant effect on the
evolution of large patches, leading to ``stochastic inflation''
\cite{Vilenkin:1983xq}.  Such fluctuations might lead to a situation
in which part or even in some sense the majority of the universe
continues to inflate for all time, i.e.\ ``chaotic eternal inflation''
\cite{Linde:1986fd}.

The advent of the ``string landscape''
\cite{Bousso:2000xa,Susskind:2003kw} with its  
complicated vacuum structure has reinvigorated the search for a
suitable measure on inflationary histories in situations where more
than one possible history can be conceived of.  Much of the debate
revolves around the extent to which predictions should be conditioned
on observations and, if more inflation leads to more observers,
whether and how any ``volume-weighting'' should be implemented.  For
technical reasons much of this recent work has focussed on models
where the inflaton is expected to ``tunnel'' from one vacuum state to
another via bubble nucleation
\cite{Garriga:2005av,Easther:2005wi,Vanchurin:2006qp,Bousso:2006ev,Aguirre:2006ak,Aguirre:2006na,Linde:2007nm,DeSimone:2008if,Linde:2008xf}.
\cite{Tegmark:2004qd} is an exception, considering random initial
conditions in random potentials, and the ``reheating-volume'' approach
has been applied to both stochastic and bubble nucleation
models~\cite{Winitzki:2008yb,Winitzki:2008ph,Winitzki:2008jp}. Quantum
cosmological studies 
\cite{Hartle:1983ai,Hawking:1998bn,Gratton:2000fj,Hawking:2002af,Hawking:2003bf,Hawking:2007vf,Hartle:2007gi,Hartle:2008ng,Hartle:2009ig,Hartle:2010vi} provide complementary perspectives.  

The approach discussed in this paper illuminates and expands the
approach to stochastic inflation and volume-weighting presented in
\cite{Gratton:2005bi}, in which one follows the evolution of the inflaton in a
$\lambda \phi^4$ potential in proper time with a Langevin noise term
approximating the quantum fluctuations.  There expectation values were
calculated for the field history and perturbatively corrected for
the effects of volume weighting.  By allowing for final-time
constraints on the field value and considering weighting field values
at some time by either the volume at that time or the volume at
the final time, \cite{Gratton:2005bi} began to directly attack the two
issues in 
the debate mentioned above.  The current paper addresses more general
inflationary models than $\lambda \phi^4$ and in some sense corrects
the perturbative 
conclusions of \cite{Gratton:2005bi} via a non-perturbative treatment
of volume 
weighting by way of a path integral.  The change in viewpoint is
similar to that in going from 
the Heisenberg approach to the Feynman approach in quantum mechanics
when trying to address a question about the history of the system.
An early approach to a Langevin model of stochastic inflation was
presented by Hodges in~\cite{Hodges:1989zz}; more recent work includes
\cite{Martin:2005ir,Martin:2005hb}.
Refs.\ \cite{Creminelli:2008es,Dubovsky:2008rf} also address eternal
inflation 
in a related manner.  In contrast, much of the early work on stochastic eternal
inflation
\cite{Goncharov:1987ir,Nakao:1988yi,Sasaki:1988df,Sasaki:1987gy,Linde:1993xx}
 attempted to follow 
  in time the evolution of a probability 
distribution for the inflaton with a Fokker-Planck equation (analogous
to the Schrodinger approach to quantum mechanics).  Such approaches
typically broke down after a finite time, when the probability
became unnormalizable rising
rapidly with field value, leading to the suspicion that Planck-scale effects
might be vital in controlling the theory and restoring predictivity.
This led in part to proper time volume weighting falling out of favour
as a measure on eternal inflation.  In addition, puzzles such as the
``Youngness Paradox'' \cite{Guth:2007ng} (\textemdash if a fraction
more inflation produces exponentially 
more volume, aren't the most common observers at a given time the
youngest ones conceivable?\textemdash) seem particularly acute with
proper time volume 
weighting.  We will see the surprising way in which a constrained
path integral approach mitigates all these issues and so it may be suggested
that proper time volume weighting should be reinstalled as a useful
measure for at least some calculations in stochatic inflation.

This paper is organised as follows.  First, a measure on slow-roll
inflationary histories is presented.  Saddle points of histories are
discussed, and then volume-weighting is introduced.  The $\lambda
\phi^4$ model is studied in depth, and the way inflation typically
ends is investigated.  Finally there is a discussion and conclusions.  

\section{\labsec{measure}A Measure on Slow-Roll Histories}

In this section we derive a measure on slow-roll inflationary
histories, starting from the appropriate Langevin equation for
slow-roll inflation (see e.g.\
\cite{Gratton:2005bi}; note that for comparison with other works a
$3/(2\pi)$ factor omitted there has been restored here):   
\ba
\dot{\phi}+\frac{V_{,\phi}}{3H}=\frac{1}{2\pi} H^{3/2} n(t)
\labeq{langevin}
\ea
where $n(t)$ is a Gaussian-normalized white noise term and
\ba
H=H(\phi)=\sqrt{V(\phi)/(3M^2)},
\labeq{friedmann}
\ea with $M$ being the reduced Planck mass (henceforth assumed to be
unity).  One might think of  
Eq.\ \refeq{langevin} as describing the evolution of a member of an
ensemble of physical-Hubble-volume sized regions forward in time.
Note that 
\refeq{friedmann} determines $H$ as a function of $\phi$, and thus the
scale factor 
$a$ at a time $t$ as a function of the history of $\phi$ up to time $t$.

Now consider an arbitrary history $\phi(t)$.  After a short time
$\Delta t$, the field will be at $\phi(t)+\dot{\phi}(t) \Delta t$, a
change of $\Delta \phi = \dot{\phi}(t) \Delta t$.  From
\refeq{langevin}, the change of the field should be centered on 
$-V_{,\phi} \Delta t/ (3H)$
with a variance of $H^3 \Delta t/(4 \pi^2)$.
So the probability of this segment of history occurring is:
\ba
\sqrt{\frac{2\pi} {H^3 \Delta t}}
  e^{-2 \pi^2\(\dot{\phi}+\frac{V_{,\phi}}{3H}\)^2 \Delta t/ H^3}.
\labeq{probseg}
\ea
Multiplying to obtain the joint probability for the entire history,
and taking the limit $\Delta t \rar 0$, we obtain:
\ba
P[\phi(t)] D\phi \sim  e^{-\int_0^T L_0(\phi) dt} D\phi
\labeq{phihistmeas}
\ea
with a ``Euclidean Lagrangian''
\ba
L_0=\frac{2\pi^2 \( \dot{\phi}+\frac{V_{,\phi}}{3H}\)^2}{H^3}.
\labeq{euclang}
\ea
Furthermore, inspection of \refeq{probseg} suggests a change of
variable that both renders the prefactor in \refeq{phihistmeas}
independent of field and 
makes the kinetic term in \refeq{euclang} canonical (up to a surface term):  
\ba
q \equiv \int \frac{2\pi}{H^{3/2}} d\phi
\labeq{qdef}
\ea
leading to
\ba
P[q(t)] Dq = e^{-\int_0^T L_0(q) dt} Dq
\labeq{normalmeasure}
\ea
up to a numerical factor.  Here 
\ba
L_0 (q)=\frac{1}{2}(\dot{q}+f(q))^2
\ea
with
\ba
f \equiv  \frac{2\pi V_{,\phi}}{3 H^{5/2}}
\ea
expressed in terms of $q$.\footnote{An alternate derivation following
  the lines of \cite{Parisi:1979ka} can yield a determinant correction to
  the measure coming from the change of variables from noise
  realizations to field realizations.  For the $q$ variable for a
  $\lambda \phi^4$ potential, this determinant is independent of field
  so both derivations agree precisely.  Given the already heuristic nature of
  our starting point, Eq.~\refeq{langevin}, we do not consider such
  corrections further in this work.}

\section{\labsec{pi}The Path Integral and its Saddle Points}

Given a general measure $e^{-S[q(t)]} Dq$ on histories $q(t)$ along with
a specification of the class of histories to integrate 
over, one may
calculate the
expectation value of some quantity of interest, $A$ say, with a path integral:
\ba
\< A \> =\frac{\int Dq \, A \, e^{-S[q(t)]}}
{\int Dq \, e^{-S[q(t)]}}.
\labeq{exval}
\ea
Note that $A$ can
be of a very general nature, either local or non-local in
time for example.  

As in the Feynman path integral approach to quantum
mechanics, it is often useful to look for saddle points in
approximating \refeq{exval}.  
Corresponding to \refeq{normalmeasure} for example we would take
$S[q(t)]=S_0 [ q(t)]\equiv\int_0^T 
L_0(q)dt$, and, certainly in this case, finding saddle points is
very simple, since the lagrangian is equivalent to that for a point
mass moving in some potential.  There even exists a conserved energy.
Furthermore, we can work with either
$q$ or $\phi$, the saddle points in either variable being equivalent.   

Once we have the saddle point solution for given initial and final
data, we can use it to approximate the probability density for that
final data given the initial data by integrating \refeq{normalmeasure}
in a gaussian approximation.  The leading term is simply the exponential
of minus the action evaluated for the saddle point.  
By varying the field value at the final time $T$ and recalculating the
saddle point solution and the (approximate) probability density, we
can build up a picture of the probability distribution function at the
final time.  By repeating the procedure for different final times, we
can build up a picture of the evolution of the probability
distribution function in time.  For \refeq{normalmeasure}, we thus
find an approximate solution to the Fokker-Planck equation
corresponding to \refeq{langevin}.

We can also calculate the expectation value of $A$ in the
saddle point approximation. 
For a fixed final field value, we just evaluate the quantity for the
related saddle point solution.  To relax the final field value
constraint, we compute the average of $A$ calculated for all final
field values, weighted by the exponential of minus the 
action (possibly multiplied by the gaussian prefactor for higher
accuracy) for the corresponding final field value.

Let us quickly derive some useful results.  Assuming the action $S$ is
expressible as an integral over a 
local-in-time Lagrangian $L$, then we can introduce a momentum 
\ba
p_q \equiv \frac{\partial L}{\partial \dot{q}}
\ea
and Hamiltonian $H'=p_q \dot{q}-L$ (primed to avoid confusion with the
Hubble parameter $H$) and corresponding energy $E$.  Just as in
classical mechanics 
(see e.g.\ \cite{Park:1990}),
evaluating the action as a function of end point and end time, we have
$\delta S= p_q \delta q - H' \delta t$.  So, at a fixed final time $T$, the
action is extremized for a solution which both obeys the equations of
motion and has $p_q(T)=0$.  This, neglecting prefactor corrections,
gives the extremum
in the probability distribution function for $q$ at time
$T$.\footnote{Note that here, unlike in
most classical mechanics applications, one does have to be
careful not to discard total integrals, which, while not altering the
saddle point solution, affect the total value for the action and so
the probability for a history.}   As long as $L-L_0$ does not involve
$\dot{q}$, $p_q(T)=0$ implies $\dot{\phi}=-V_{,\phi}/ 3H$ at time $T$,
i.e.\ a trajectory corresponding to an extremum in the probability
distribution at time $T$ is slow-rolling as $T$ is approached.  This does not
necessarily imply that the trajectory slow-rolls all the way from $0$
to $T$, nor that extrema of the probability distribution evolve
according to slow-roll as $T$ is varied.  However, in the simple case
when $L$ is just $L_0$, both of these statements do in fact hold.   A
helpful way to verify such statements is to look at the (conserved) energy
associated with each saddle point solution.  For $L_0$, this is just 
\ba
E_0=\frac{1}{2} \dot{q}^2-\frac{1}{2} q^2
\ea
and for a solution that slow-rolls at the end time $T$, we see $E_0=0$.  As
$E_0$ is conserved along the saddle point, we see that $\dot{q}$ must
equal $q$ back along the entire trajectory to $q=q(\phi_0)$ at $t=0$,
i.e.\ the saddle point is just the slow-roll solution evolved from the
initial condition to the time $T$.  Hence as $T$ changes the position of the
peak also follows the slow-roll solution in this case.   
    
\section{\labsec{vwpi}The Volume-Weighted Path Integral}

Now we volume-weight.  We do this by ``reweighting'' the $e^{-S_0
  [q(t)]} Dq$ measure from above by an appropriate term, and then
renormalizing.  We might imagine the physical-Hubble-volume sized
regions followed above as being ``probes'' of much larger volumes of
space that are inflating and so expanding in time.  Thus the term is
typically
 just the ``final'' volume\footnote{One could also for example
  consider weighting by 
  the volume at an intermediate time for calculating a
  final-field-value-constrained
  ``rolling'' volume average.}   
\ba
a^3 (T)=e^{\int_0^T 3H(q(t)) dt}.
\labeq{volfac}
\ea 
Because of the local-in-time nature
  of the exponent, we see that such volume weighting can be
  incorporated very simply by thinking of $S$ as an integral
  over $t$ of a more general ``Euclidean Lagrangian''
\ba
L_V = L_0 - 3 H.
\labeq{eucvollang}
\ea
We have only had to add a local-in-time
term to the potential for $q$.   This extra term alters the
constrained histories relative to the non-volume-weighted ones with
the same boundary conditions.

The volume-weighting term, only involving the field and not its
derivative, will not affect the expression for the momentum in terms
of the field and its derivative.  So, as for the
non-volume-weighted case discussed
above, the most probable trajectory will obey the slow-roll condition
at the very end.  

However, unlike before, this trajectory will not
have slow-rolled all the way from the initial condition; there are
two additional effects that cannot generally cancel.  First, the equation of
motion now has an extra field-dependent term.  Second, the expression
for the energy $E$ has an additional $+3H(\phi)$ term, so, evaluating this
at the end of the trajectory, the energy of
the solution is moved from zero to $E=3 H (\phi(T))$.    Thinking
momentarily of $T$ as a function of $q$ at time $T$ for the most
probable solution, we have
\ba
T=\int_{q(0)}^{q} dq' \sqrt{2\(E(q)-V(q')\)}.
\ea
Changing $E$ and $V$ as discussed to incorporate volume-weighting will
change $T(q)$ and so $q$ considered as a 
function of $T$: the peak of the probability
distribution function does not now follow slow-roll. 

We can now ask whether or not inflation ends in the rolling-volume-weighted
average.  If it does, the peak of 
the probability distribution function should pass through a field value
corresponding to a small value of $H$, i.e.\ the conserved energy should be
able to have a small value.  Evaluating the energy, expressed in terms
of $\phi$, at the initial condition thus leads to the condition:
\ba
\frac{2\pi^2 {\dot{\phi}}^2 (0)}{H^3}-\frac{2\pi^2}{H^3}
\frac{V_{,\phi}^2}{9 H^2} + 3 H \approx 0
\ea
for inflation to end in the volume-weighted average.  Now, whatever
the value of $\dot{\phi}$, if the potential term contribution to the
left hand side is positive then inflation cannot end.  So if 
\ba
H^6 > \frac{2 \pi^2 V_{,\phi}^2}{27}
\labeq{etinfcond}
\ea
then inflation cannot end in the volume-weighted average
(c.f.\cite{Creminelli:2008es}).   This precise constraint is
consistent with the 
qualitative arguments of e.g.\ Guth~\cite{Guth:2007ng} comparing the classical
movement of the 
field to the quantum fluctuations in the field over a Hubble time.
We may say that inflation is indeed ``eternal'' if the field starts at
a value such that \refeq{etinfcond} is satisfied.  Note that for $\lambda
\phi^4$ this requires $ \phi > (32 \pi^2 / \lambda)^{1/6}$.  

Let us assume that the field starts above the eternal inflation
threshold and ask what happens.  Does the volume-weighted field
asymptote to some constant value, and if so can this value be above or
below the starting value?  Or does the field average run away to a
place of infinite energy density, in either finite or infinite time?
We are minded here of the early results looking at volume-weighted
eternal inflation by solving the Fokker-Planck equation for say a
$\lambda \phi^4$ potential; there it seemed that the probability
distribution lost its extrema after a finite amount of time, becoming
unnormalizably peaked at an infinite field value.  This led to the
imposition of arbitrary boundary conditions at the Planck density and
the view that the volume-weighted field would quickly tend to its
largest possible value and stay there inflating at the maximum
possible rate.  We indeed find, investigating $\lambda \phi^4$ as a
specific example as discussed in the following section, that the
average can stop existing after a finite time.  Above we showed that
the paths corresponding to extrema of the probability distribution
must end in slow-roll. This applies to both maxima and minima, and a
minimum in the probability distribution will delimit a formal
unbounded rise in probability towards very high energy from a physical
region of field values with its own maximum.  As time goes on, the
maximum and minimum merge; the probability distribution steadily
increases with field energy.  To see whether trajectories with
Planckian energy densities are important or not for the disappearance
of the average, we now though can look at the critical saddle point
history and see whether or not it approaches the Planck density at any
stage.  For small coupling it turns out it does not and so we conclude
that Planck scale effects will not affect the disappearance of
volume-weighted average.

One may be concerned that the disappearance of the average is
indicating a failure with the whole approach.  However, we can
continue to find constrained paths to lower field values, at least for
a finite window of larger time intervals.  So perhaps the correct
interpretation is simply that answering the question ``what is the
field average on a surface of constant proper time?'' is becoming
problematical.  But because we can still answer other questions, about
constrained paths say, proper time volume weighting itself is not
obviously failing at this stage.

Pushing to larger time intervals still, we find for $\lambda \phi^4$
at least that real histories connecting the inflationary initial
conditions to low field values cease to exist.  Surely even
constrained proper time volume weighting is failing now?  This is not
necessarily the case because \emph{complex} histories now emerge that
link the initial and final conditions.  Further, as we will see below,
these complex paths are very close to being real towards the end of
inflation, and indeed basically become ``classical'' slow-roll
trajectories, insensitive to the initial field value or indeed the
time interval between the initial and final conditions.

\section{\labsec{phi4}$\lambda \phi^4$ in Depth: Complex Histories,
  Initial Conditions and the End of Inflation}  

For $\lambda \phi^4$, $q$ is proportional to $1/\phi^2$, which is in
turn proportional to the Hubble radius which we denote here by $R$ and
work with in 
order to allow for easy comparison with \cite{Gratton:2005bi}.  Introducing the
(dimensionless) constants $\alpha=8 \sqrt{\lambda/3}$ and
$\beta=\sqrt[4]{\lambda/3}/\pi$, we find 
\ba
L_V = \frac{ \( \dot{R}-\alpha R \)^2}{2 \beta^2}-\frac{3}{R}.
\ea
Saddle-point histories satisfy
\ba
\ddot{R}=\alpha^2 R+\frac{3 \beta^2}{R^2} 
\labeq{reom}
\ea
with a conserved energy
\ba
E=\frac{1}{2 \beta^2} \dot{R}^2-\frac{\alpha^2}{2\beta^2} R^2
+\frac{3}{R}. \labeq{renergy}
\ea
See Fig.~\reffig{potplot} for the associated effective potential that the
$R$ variable feels.   (Note
that in all plots $R$
and $t$ have been rescaled to absorb $\alpha$ and $\beta$ via $t \rar
\alpha t$, $R \rar (\alpha/\beta)^{(2/3)} R$.)  
\begin{figure}
\includegraphics[width=8cm]{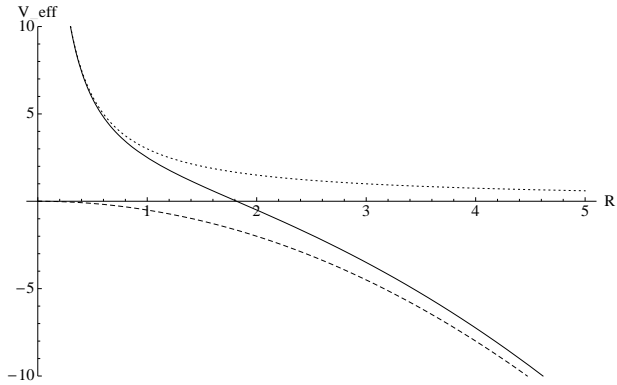}
\caption{
\labfig{potplot}
Plot of the effective potential that the $R$ variable moves in.  
}
\end{figure}

As discussed in Sec.~\refsec{vwpi}, the momentum 
\ba 
p_R = \frac{1}{\beta^2} \(\dot{R}-\alpha R \) 
\labeq{rmom} 
\ea 
has to be
zero at the end of a history that finishes at an extremum of the
probability distribution function at time $T$.  Hence, from
\refeq{renergy}, the energy of such a path is $3 / R(T)$, which is
small for weak coupling (of order $\lambda^{-1/2}$) if $R(T)$
corresponds to $\phi \sim 1 $ for which inflation would classically be
ending. Eternal inflation is inevitable if $R$ ever becomes less than
$\sqrt[3]{6 \beta^2 / \alpha^2}$ (the zero of the effective potential
in Fig.~\reffig{potplot}).

Fig.~\reffig{potplot} is very helpful for understanding the nature of the
(real) saddle point solutions, illustrating the discussion of
Sec.~\refsec{vwpi}. 
Let us look for solutions connecting $R=\ri$ to $R=\rf$
  (``i'' for initial, ``f'' for final). If $\rf > \ri$, there are two
  classes of solutions differentiated by the sign of the initial velocity.
  One rolls up the potential, turns round and rolls back down past \ri\ on
  the way to \rf; the other rolls straight down the potential from \ri\
  to \rf.  If $\rf < \ri$,
  there are again two classes of solution, now differentiated by the
  sign of the final velocity.  In one, $R$ rolls straight
  up the potential from \ri\ to \rf, in the other, $R$ rolls up the
  potential passing through \rf, then rolls back down to \rf.  If
  $\rf=\ri$, there is only one class of solution, $R$ rolling up the
  potential and then back down. In all
  cases, varying the initial speed, or equivalently the energy $E$,
  changes the time $T$ needed to go from \ri\ to \rf.  

Scanning over \rf, $E$ and the velocity sign and recording the
time $T$ each solution takes gives us complete information about the
behaviour of the probability distribution function for $R$ as a
function of $T$.   Note though that there is no guarantee that
arbitrarily large values of $T$ will be obtained in the scan, and
indeed $T$ turns out to be bounded when volume weighting is switched on.   

If we temporarily focus on the subset of histories with zero final
momentum (and 
so with energy $E=3/\rf$ with volume-weighting, or with $E=0$ without),
corresponding to the extrema of the 
probability distribution function, we can build up a sketch of the
loci of the maxima and minima of the probability distribution function
as a function of time as in Fig.~\reffig{mmloc}. 
\begin{figure}
\includegraphics[width=8cm]{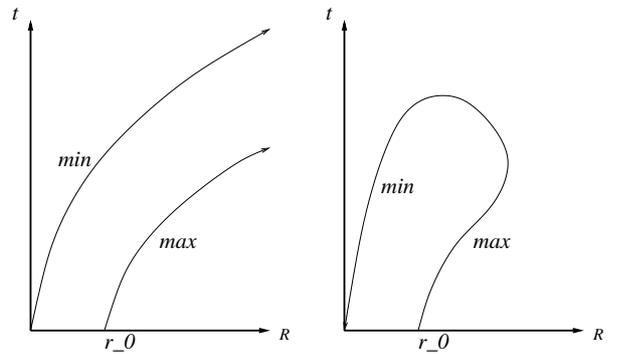}
\caption{
\labfig{mmloc}
Sketch of the loci of the minima and maxima of the volume-weighted probability
distribution function for $R$ as a function of time.  The left panel
is for starting
below the eternal inflation threshold, while the right panel is for
starting above the eternal inflation threshold.  (Without
volume-weighting the plot would be qualitatively similar to the
``max'' branch of the left hand panel.)
}
\end{figure}
Without volume-weighting, a solution exists  for any $T$ (the field spending
longer and longer at small $R$ as $T$ increases), with \rf\ always
greater than \ri, and the distribution
moves (exponentially in time) to larger $R$ as time passes.  Switching on
volume weighting corresponds to adding a repulsive force, requiring $R$ to
start with a more negative velocity than for 
the non-weighted 
case with the same \rf\ and $T$.  Thus $R$ gets smaller more rapidly
as expected; 
volume-weighting favours higher field values.  Starting well below the
eternal inflation threshold, the picture 
is qualitatively similar to the non-volume-weighted case.  Starting
above the eternal inflation 
threshold, the picture changes significantly however.  The steep,
``brick-wall'', nature of the repulse volume-weighting 
term in the effective potential means that there is in fact an upper limit on
how much $T$ can be increased by increasing the initial speed of $R$.
After this time there are no extrema; the probability density
increases monotonically towards small $R$.   We are able to conclude
that at some intermediate time, at some $R> \ri$, the peak of the
probability distribution turns around; the inflaton begins to climb
back up its potential in the volume-weighted average.  The largest
value of \rf\ attained can be deduced by equating the effective
potential at $R=\ri$ to the energy $E=3/\rf$ evaluated at the end.

Returning to the general case, it may seem strange that constrained
saddle point solutions linking 
\ri\ to \rf\ fail to exist for too large of a time difference.  As
mentioned above, the paradox is resolved by realizing that there is 
no necessity for the saddle point histories to be real.  Just as in
contour integration, where one may deform real contours into the complex
plane to pass through a saddle point in order to apply the method of
steepest descents, 
we can do likewise here.  Indeed, the use of complex histories has
a precedent in the 
Euclidean ``No-Bounday'' approach to quantum cosmology, where for
``large'' final three-geometries the Euclidean path integral has a
complex saddle point with a Lorentzian part \cite{Hartle:1983ai}.  We
need only preserve our boundary conditions, namely that $R(0)=\ri$ and
that $R(T)=\rf$.  We see straight away that the imaginary
component of $R$ has to be zero at both ends, but that there is no
such constraint on the imaginary component of $\dot{R}$ at the ends.
Decomposing 
\refeq{reom} into real and imaginary parts, we can visualize $R$ as a
point
moving in a two-dimensional force field as heuristically plotted in
Fig.~\reffig{forceplane}.  
\begin{figure}
\includegraphics[width=8cm]{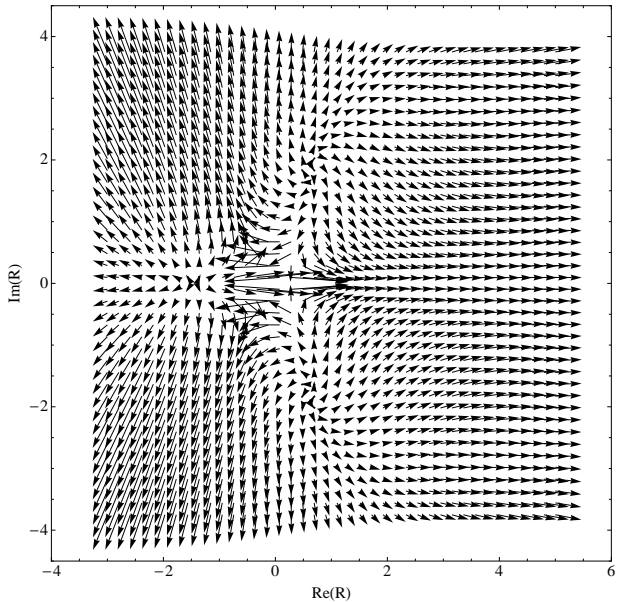}
\caption{
\labfig{forceplane}
Heuristic plot of the force field determining the motion of $R$ in the
complex $R$-plane. Note the zero-force ``loitering points'' at the
solutions of $R^3 =-3$.  Planckian energy density, $\lambda \phi^4
\sim 1$, corresponds to the
rescaled 
$R$ shown being small, of order $\lambda^{1/6}$ for weak coupling.  An
eternal-inflationary starting point would be along the real axis
between $0$ and $\sqrt[3]{6}$.  The 
end of inflation, $\phi \sim 1$, corresponds to the rescaled $R$ shown
becoming large and 
positive, of order $\lambda^{-1/3}$.   
}
\end{figure}
We note the (unstable) zero-force locations specified by $R^3= - 3 \beta^2 /
\alpha^2$.  By tuning the solution so that it approaches one of these
points with near zero speed, it is possible for $R$ to ``loiter''
there for as long as is needed.  Solutions with a long loitering
period must have a complex energy very close to that of the effective
potential evaluated at the loitering point in question.  This
complex energy determines solutions that asymptotically reach the
loitering point in the future from \ri\ or from the past from \rf.
Appropriate deviations in the initial velocities will ``connect up'' the
two asymptotic solutions and make the total solution last for the
desired finite time $T$.

For solutions linking inflaton values corresponding to starting above
the eternal-inflationary threshold to field values towards the end of
conventional inflation, the appropriate loitering points are the ones
with a positive real component and non-zero imaginary component of
$R$.  These will provide a conjugate pair of histories.  Each history
will approach its respective loitering point and then roll back
towards the real axis out to large positive values of $R$.

Focusing on a single member of a conjugate pair for clarity, the way
the solution reaches \rf\ will become almost independent of how large
$T$ is, as long as $T$ is sufficiently large.  We thus see in a
precise way how eternal inflation ``loses memory'' of initial
conditions, in that, at sufficiently late times, the way inflation typically
ends is very insensitive to the initial field value.

Note that for weak coupling the history need not go particularly close
to $R \lesssim 1$ and the Planckian energy densities there.  Rather,
the amplitude of the loitering points corresponds to $\phi \sim
\lambda^{-1/6}$, of order the eternal inflation threshold.  Hence
conclusions drawn from the history may hoped to be insensitive to any
Planck-scale corrections to the model.

The imaginary part of a trajectory corresponding to the late-time end
of eternal inflation for weak coupling is very small, and the real
part of the field 
basically slow-rolls, as illustrated in Fig.~\reffig{solndiff}.  
\begin{figure}
\includegraphics[width=8cm]{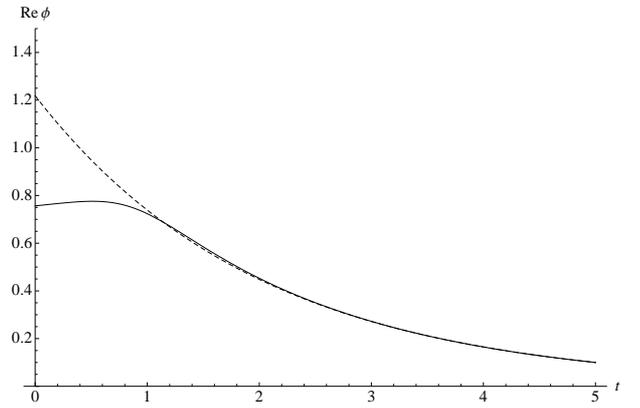}
\caption{
\labfig{solndiff}
Plot of the loitering solution (solid line) departing from the
slow-roll solution (dashed line) to the past of a region where
inflation ends. The difference only becomes significant when the field
approaches the eternal inflationary regime; the end of inflation is
classical.} 
\end{figure}
These
statements can be made quantitative by rearranging \refeq{renergy} to
express $\dot{R}$ in terms of $R$ and $E$,
\ba
\dot{R}= \alpha R \sqrt{1+ \frac{2 E- \frac{6 \beta^2}{R}}{\alpha^2
    R^2}},
\ea  
the fractional correction over the slow-roll result ($\dot{R}=\alpha R$) being
small when the modulus of $R$ is well away from the eternal
inflationary regime.

We have just seen that where stochastic eternal inflation ends, it
basically ends classically.  This might appear counter-intuitive,
given the Youngness Paradox arguments about proper-time volume
weighting.  Consider looking for regions where the inflaton, if it
slow-rolled, would have either one or two efolds say left to go.  The
Youngness Paradox would suggest that there are exponentially many more
regions with one efold left than regions with two efolds left, the
latter histories being ``younger'' and so having had more inflation in
their past.  One might have also thought that to the past of any of
these regions the field would be much higher up its potential than
slow-roll would suggest, perhaps even up at Planck-density values,
since such histories would have exponentially more volume.  Our result
is not inconsistent with the first conclusion but suggests that
volume-weighting does not sufficiently dominate over classical motion
for the second conclusion to apply also.  Thus standard calculations
of inflationary density perturbations are probably safe even in
eternal inflation as long as the coupling is weak.

\section{Discussion and Conclusions}

We can assemble what we have learned above into a somewhat cogent picture
of volume-weighted stochastic eternal inflation.  The field must start
off above the eternal inflationary threshold, and then we soon see the
volume-averaged field stop decreasing and turn around and begin
increasing, indicating that volume effects are outweighing classical
drift.  From this we may hope that a late-time ``steady state''
situation will arise with late-time results dominating any averaging.
After a finite proper time, the volume-averaged field ceases to exist;
the system is dominated by strong fluctuations and a ``global''
picture breaks down.  Nontheless, we may choose to focus on the
observationally relevant but rare regions of the universe where
inflation happens to end.  Then we find that inflation ends in
practically the same slow-roll manner on all proper time slices and
hence some level of predictability is restored.  The saddle point
histories deviate into the complex plane rather than continue to
values far above the eternal inflation threshold, indicating that the
conventional view of the inflaton as jumping up and down on its potential
in the eternal inflation regime might be too simplistic. Because when
inflation does end it basically ends in slow-roll, conventional
density perturbation calculations should still apply, preserving the
successful predictions of conventional treatments of inflation.

The general techniques and insights of this paper should apply to many
large-field models of inflation.  It would be interesting to
investigate potentials with multiple vacua.  Indeed, for
``Mexican-hat'' type potentials, $V =\lambda (\phi^2-\phi_0^2)^2$, one
can analytically obtain an expression for $q$ in terms of $\phi$ and
so obtain an explicit formula for the effective potential for $q$.
Thus one could investigate volume-weighting for small-field models of
inflation where would might expect its effect to be less pronounced
than for the large-field case studied here.  (Note that an early
work~\cite{Rey:1986zk} discusses an approximate path integral
treatment of the behaviour of the inflaton in a ``new'' inflationary
potential.)  Numerical investigation of the determinant prefactor
would be helpful in getting an idea in how ``classical'' the histories
really are.  The author has checked numerically that there are no
negative modes satisfying the relevant boundary conditions for a
sample of representative (real) histories, as one would hope.  It
would also be possible to go beyond slow-roll, obtaining
fourth-order equations for the saddle point histories, though the
precise way in which the quantum fluctuations are modelled might need to be
thought through more carefully.

As in quantum mechanics, we have seen that a path integral approach is
particularly useful when asking time-dependent questions and looking
for semi-classical histories.  It has given us a technique for calculating in
volume-weighted eternal inflation that is relevant for observations.  We
have been able to demonstrate how inflation typically ends normally
even with volume-weighting in a manner insensitive to the precise initial
conditions.  By retreating from demanding a global picture of the
universe at all times and rather adopting a more ``top-down'' observationally
relevant approach \cite{Hawking:2002af,Hawking:2003bf,Gratton:2005bi}
the path integral has allowed us to push much further 
than in the Fokker-Planck approach without having to worry about
Planck density issues.  We have also obtained a different result about
the behaviour of the volume-weighted average than in the Langevin
treatment of \cite{Gratton:2005bi}.  This is possibly because that
work only perturbatively expanded around the classical solution,
implicitly forcing one to consider only the subset of histories in which
inflation has to end.

Finally, let us return to the question of proper-time volume weighting
itself.  Rather than any intrinsic flaw in the scheme, perhaps it was
the gauge-dependence of the questions that proper time volume
weighting encouraged one to ask that led to the weighting getting a
bad reputation (a canonical example of such a gauge-dependent question
being ``which value of the inflaton is most likely at a given
time?'').  A question that we have addressed in this paper is ``how
does inflation end at a given proper time?''.  Seeing that the answer
only depends very weakly on what that time actually is, we have been
able to obtain a satisfactory answer to the more general reasonable
question ``how does inflation end?'' even using proper-time volume
weighting.  So for at least some physically relevant questions perhaps
proper-time volume weighting is not so bad after all.

\begin{acknowledgments}

We thank Anthony Aguirre, Tom Banks, Paolo Creminelli, Jaume Garriga,
Alan Guth, Thomas Hertog, Antony Lewis, James Martin, Paul Steinhardt,
Neil Turok and Toby Wiseman for helpful discussions.  This research
was supported in part by a minigrant from The Foundational Questions
Institute (fqxi.org).  I am supported by STFC.

\end{acknowledgments}

\bibliography{volweight.bib}

\end{document}